\documentclass[spanish,english]{revtex4}
\usepackage[T1]{fontenc}
\usepackage[latin9]{inputenc}
\usepackage{textcomp}
\usepackage{amsmath}
\usepackage{amssymb}
\usepackage{graphicx}
\usepackage{esint}

\makeatletter
\@ifundefined{textcolor}{}
{%
 \definecolor{BLACK}{gray}{0}
 \definecolor{WHITE}{gray}{1}
 \definecolor{RED}{rgb}{1,0,0}
 \definecolor{GREEN}{rgb}{0,1,0}
 \definecolor{BLUE}{rgb}{0,0,1}
 \definecolor{CYAN}{cmyk}{1,0,0,0}
 \definecolor{MAGENTA}{cmyk}{0,1,0,0}
 \definecolor{YELLOW}{cmyk}{0,0,1,0}
}

\makeatother

\usepackage{babel}
\addto\shorthandsspanish{\spanishdeactivate{~<>}}

\begin{document}
\selectlanguage{spanish}%
\global\long\def\abs#1{\left| #1 \right| }
\global\long\def\ket#1{\left| #1 \right\rangle }
\global\long\def\bra#1{\left\langle #1 \right| }
\global\long\def\half{\frac{1}{2}}
\global\long\def\partder#1#2{\frac{\partial#1}{\partial#2}}
\global\long\def\comm#1#2{\left[ #1 ,#2 \right] }
\global\long\def\vp{\vec{p}}
\global\long\def\vpp{\vec{p}\, ^{\prime}}
\global\long\def\dt#1{\delta^{(3)}(#1 )}
\global\long\def\Tr#1{\textrm{Tr}\left\{  #1 \right\}  }
\global\long\def\Real#1{\mathrm{Re}\left\{  #1 \right\}  }
\global\long\def\braket#1{\langle#1\rangle}

\selectlanguage{english}%

\title{Wigner formalism for a particle on an infinite lattice: dynamics
and spin}

\author{M Hinarejos$^{1}$, M C Bañuls$^{2}$, and A Pérez$^{1}$ }

\address{$^{1}$Departament de Física Teòrica and IFIC, Universitat de València-CSIC,
Dr. Moliner 50, 46100-Burjassot, Spain\\
$\text{\texttwosuperior}$Max-Planck-Institut für Quantenoptik, Hans-Kopfermann-Str.
1, Garching, D-85748, Germany.}
\begin{abstract}
The recently proposed Wigner function for a particle in an infinite
lattice {[}NJP 14, 103009 (2012){]} is extended here to include an
internal degree of freedom, as spin. This extension is made by introducing
a Wigner matrix. The formalism is developed to account for dynamical
processes, with or without decoherence. We show explicit solutions
for the case of Hamiltonian evolution under a position-dependent potential,
and for evolution governed by a master equation under some simple
models of decoherence, for which the Wigner matrix formalism is well
suited. Discrete processes are also discussed. Finally we discuss
the possibility of introducing a negativity concept for the Wigner
function in the case in which the spin degree of freedom is included.
\end{abstract}
\maketitle

\section{Introduction}

Since its introduction, the Wigner function (WF) \cite{PhysRev.40.749}
has played an important role in physics. Quantum mechanics can be
entirely formulated using this tool, therefore providing an alternative
description of quantum phenomena, along with their dynamics. Also
from a more experimental perspective, the WF has proven instrumental
for tomographic reconstruction of the states prepared in the lab.
The WF is in hence completely equivalent to the standard quantum mechanical
formalism. Nevertheless, the particular features of the phase space
description make it advantageous in some situations, for instance
recognizing the quantum features of states, or dealing with decoherence
scenarios. In the WF interference effects manifest in a clear way
\cite{Lee2011,PhysRevA.74.042323,Hillery1984121,citeulike:4313181}.
Another interesting property that manifests in the visualization of
the WF of some states is the appearance of negative values over the
phase space. This fact has been considered as a direct manifestation
of the quantum nature of such states, and used to characterize their
\textit{quantumness} \cite{kenfack,mari12,PhysRevLett.106.010403}.
The relativistic extension of the Wigner function \cite{PhysRevD.13.950}
has also found applications to a wide variety of problems, ranging
from general relativistic kinetic theory and statistical mechanics
\cite{nla.cat-vn1394639,Hakim:1379544}, nuclear matter at high densities
and temperatures \cite{DiazAlonso:1991hy}, electrons in magnetic
fields \cite{Hakim1982,Yuan2010}, the quark-gluon plasma \cite{Elze1986},
to neutrino propagation in astrophysical or cosmological scenarios
\cite{Sirera:1998ia,YAMAMOTO2003}. 

The applications mentioned above make use of a Wigner function defined
in continuous space. It is nevertheless possible to introduce also
a sensible Wigner function for systems on a discrete space. The definition
for the case of a finite dimensional Hilbert space can be traced back
to Stratonovich and Agarwal \cite{stratonovich,PhysRevA.24.2889}
(see also \cite{Varilly1989}), who introduced a spherical, continuous
phase space for a spin particle. A possible generalization was proposed
by Wootters in 1987~\cite{Wootters1987} for prime dimensional systems,
and later generalized to any power of primes in~\cite{PhysRevA.70.062101}.
A different construction was followed in~\cite{leonhardt95prl,PhysRevA.53.2998,miquel02qc}
which could cope with any dimension of the Hilbert space at the expense
of enlarging the size of the phase space grid (see \cite{vourdas04review,ferrie11review}
for a review). The discrete WF for a finite dimensional system is
furthermore related to quantum information problems \cite{galvao05speedup,ferrie11review,Bianucci2002,miquel02qc,PhysRevA.72.012309,veitch12,mari12}. 

If the discrete Hilbert space is infinite dimensional, a different
extension of the WF is required. In \cite{hinarejos12} we proposed
a definition of the WF that can be used for such systems, having the
correct marginal properties and with the advantage that a closed form
can be obtain in some cases, such as the Gaussian states. Notice that,
in contrast to the continuous case, where the axiomatic definition
of the WF uniquely determines its functional form \cite{bertrand87},
in the discrete case different definitions are possible that respect
the mathematical conditions enumerated above (see also \cite{PhysRevA.49.3255,rigas11}
for alternative, related definitions, motivated by the study of the
angle and angular momentum phase space). 

Many of the problems where the continuous WF has found application
concern particles with spin, or with spinor descriptions of quantum
fields. In order to use the phase space formalism in this scenario,
a generalization has to be introduced which combines the spin and
spatial degrees of freedom (dof). One of the most common prescriptions
in the literature is the use of a matrix valued WF \cite{PhysRevA.56.1205},
where the spinor or spin indices give rise to various matrix elements.
Indeed, other possibilities exist, such as introducing a phase space
for the spin degrees of freedom, which correspond to another discrete,
finite dimensional Hilbert space, and construct a real valued WF for
the cartesian product of spin and space phase spaces. In the matrix-valued
WF, the treatment of space and spin dof is not symmetric. The spatial
part is described in terms of a phase space, while the spin is unchanged.
Although the treatment is asymmetric, such description has some advantages
when dealing with a particle subject to a spin-dependent force, since
some effects like the spin precession, or motion that depends on the
spin component, are better visualized with respect to a fixed spin
basis. Examples of this description are the analysis of the Stern-Gerlach
experiment \cite{Utz2015}, the study of entangled vibronic quantum
states of a trapped atom \cite{PhysRevA.56.1205}, or the reconstruction
of the full entangled quantum state for the cyclotron and spin degrees
of freedom of an electron in a Penning trap \cite{Massini2000}. 

In this paper, we have extended the definition of the Wigner function
introduced in \cite{hinarejos12} to incorporate the spin of a particle,
using the Wigner matrix formalism for the spin degrees of freedom,
and we illustrate the consequences of this definition by analyzing
some simple physical situations, such as states involving spatial
and spin entanglement or dynamical evolution, as it appears for a
particle subject to a spin-dependent force. 

The rest of this paper is organized as follows. In Section II we introduce
a definition for the Wigner matrix (WM) that incorporates the spin
of the particle, and summarize the main properties that are satisfied
by this object. To illustrate the structure of this representation,
we consider some simple cases in Section III. Section IV contains
the main results of our paper, concerning the dynamics obeyed by the
WM under the influence of an interacting Hamiltonian that may depend
or not on the spin. First, we study the time evolution in continuous
time, by deriving the equation of motion for the WM and solving this
equation in some simple cases. The situation without spin serves us
to consider the special case of a particle on a lattice interacting
with a linear potential. We also investigate the interaction that
appears for a spin-dependent force to visualize the main differences
with the spinless case. Finally, we study the effect of decoherence
for the system under consideration. Also in this Section, we show
how one can make use of the WM to investigate the dynamics that appear
in some discrete-time problems, and consider the particular example
of the quantum walk. As before, we show the effect that decoherence
may have on such problems.

One of the advantages of a WF description of continuous variable systems
is the access to a negativity that measures the non-classicality of
states. Although the relation of the negativity to non-classicality
is well established, this quantity does not correspond to a physical
observable. With a more general definition as the WM and the occurrence
of (non-classical) spin degrees of freedom, we may wonder if there
is a generalized negativity quantity and whether it retains some physical
information. This is discussed in Section V. Section VI presents our
main conclusions. The derivation of some formulae has been relegated
to the Appendix in order to make our presentation more transparent.

\section{Particle with spin on a one-dimensional lattice}

We are interested in the phase space description of a spin 1/2 particle
that is allowed to move on an infinite 1D lattice. A paradigmatic
example is the quantum walk (QW) on the line, where a particle moves
along the sites of a 1D lattice. In its discrete-time version \cite{Aharonov93},
the direction of motion is dictated by the state of an extra two-dimensional
Hilbert space (the \textit{coin}), that can correspond to the internal
spin of the moving particle. In fact, during the process the spatial
and internal states become entangled, even if the initial state was
separable, thus making clear the need for a joint description of both
degrees of freedom. Another example is the study of spin dependent
transport properties of single atoms in a 1D optical lattice \cite{Karski2011}.

We will start with the definition of the WF for a (spinless) particle
on a 1D lattice already introduced in \cite{hinarejos12}. We consider
a lattice with sites $\{na/n\in\mathbb{Z}\}$, where $a$ is the lattice
spacing. To these sites one can associate a basis $\{\ket n$\}, with
$n\in\mathbb{Z}$. By a Fourier transformation we define a quasi-momentum
basis, $|q\rangle=\sqrt{\frac{a}{2\pi}}\sum_{n}e^{iqna}|n\rangle$,
which can be restricted to the first Brillouin zone, $q\in[-\frac{\pi}{a},\frac{\pi}{a}[$.
The phase space is defined by points $(m,k)$, where $m\in\mathbb{Z}$,
whereas $k$ is continuous and periodic, taking values in $[-\pi,\pi[$.
With these notations, we define the WF as
\begin{equation}
W(m,k)\equiv\mathrm{tr}\left[\rho A(m,k)\right]=\frac{1}{2\pi}\sum_{n}\langle n|\rho|m-n\rangle e^{-i(2n-m)k},\label{eq:WFnospin}
\end{equation}
where $\rho$ is the density operator corresponding to the state of
the system, and $A(m,k)=\frac{1}{2\pi}\sum_{n}|m-n\rangle\langle n|e^{-i(2n-m)k}$
are the phase point operators for the lattice. It can be checked that
the above definition fulfills the necessary requirements to be considered
a valid WF. We refer the reader to the above reference for more information
about the properties obeyed by (\ref{eq:WFnospin}). 

We now would like to incorporate the additional degree of freedom
arising from the spin of the particle. As discussed in the Introduction,
there are different approaches in the literature to describe finite
dimensional Hilbert spaces, such as the spin of a particle. One can
combine both degrees of freedom (spin and lattice) by a tensor multiplication
of the corresponding point operators, as done in \cite{Luis2005}
for angular momentum and spin states. 

As discussed in the introduction, here we opt for a prescription with
ample acceptance in the continuous applications, namely a matrix-valued
WF. A similar choice has been used in relativistic and non-relativistic
setups with continuous spatial dof. Among the latter we can mention
the study of Stern-Gerlach experiment \cite{Utz2015}, the analysis
of entangled vibronic quantum states of a trapped atom \cite{PhysRevA.56.1205},
or the reconstruction of the full entangled quantum state for the
cyclotron and spin degrees of freedom of an electron in a Penning
trap \cite{Massini2000}. The Wigner function defined in this way
combines the following properties:

- It keeps a close analogy with the definition of the relativistic
Wigner function \cite{PhysRevD.13.950,nla.cat-vn1394639,Hakim:1379544},
thus allowing to describe the transition from the relativistic to
the non relativistic regime. 

- It appears as a simple and convenient choice to describe the spin
motion in some particular cases, like the Stern-Gerlach experiment
in continuous space \cite{Utz2015}, or the dynamics of a spin 1/2
particle on a lattice under the effect of a spin-dependent force,
as described in Sect. IV.

We consider the Hilbert space $\mathcal{H}=\mathcal{H}_{l}\otimes\mathcal{H}_{s}$,
where $\mathcal{H}_{l}$ stands for the motion on the lattice, and
$\mathcal{H}_{s}$ describes the spin states. The composed Hilbert
space is spanned by the basis $\{\ket{n,\alpha}\equiv\ket n\otimes\ket{\alpha}\}$
with $n\in\mathbb{Z}$ and $\ket{\alpha}/\alpha=0,1$ designate the
eigenvectors of the $\sigma_{z}$ Pauli matrix (these states might
also correspond to the computational basis of a qubit, or to the levels
of a two level system). According to the above discussion, we propose
the following definition for the WM 

\begin{equation}
W_{\alpha\beta}(m,k)\equiv\frac{1}{2\pi}\sum_{n}\langle n,\alpha|\rho|m-n,\beta\rangle e^{-i(2n-m)k}.\label{eq:WFspin}
\end{equation}
We then have a set of four functions $W_{\alpha\beta}(m,k)$, $\alpha,\beta=0,1$
forming a $2\times2$ matrix. Each function, as before, is defined
on the phase space of points $(m,k)$, with $m\in\mathbb{Z}$, and
$k$ takes values in $[-\pi,\pi[$. A similar definition can be made
for any operator $\mathcal{O}$ acting on $\mathcal{H}$:

\begin{equation}
W_{\alpha\beta}^{\mathcal{O}}(m,k)\equiv\frac{1}{2\pi}\sum_{n}\langle n,\alpha|\mathcal{O}|m-n,\beta\rangle e^{-i(2n-m)k}.\label{eq:WFspinopA}
\end{equation}
Unlike the spatial variables, where the relationship with phase space
points is non trivial, there is a direct correspondence between spin
indices in the state of the system and indices in the matrix WF. This
implies that operations on the spin space, such as rotations, change
of basis or interactions with a spin-dependent force, as studied below,
become more transparent using the matrix WF than other kind of representations
for the spin. Moreover, the definition Eq. (\ref{eq:WFspin}) keeps
a closer analogy, for pure states, to the relativistic WF used in
Quantum Field Theory. For such states one has $\rho=\ket{\Psi}\bra{\Psi}$
and we can write
\begin{equation}
W_{\alpha\beta}(m,k)\equiv\frac{1}{2\pi}\sum_{n}\Psi_{\alpha}(n)\Psi_{\beta}^{*}(m-n)e^{-i(2n-m)k}\label{eq:WFspinors}
\end{equation}

with $\Psi_{\alpha}(n)\equiv\langle n,\alpha\ket{\Psi}$. In the continuum
limit, the functions $\Psi_{\alpha}(n)$ can be interpreted as the
components of a Pauli spinor or a Dirac spinor. In this case, Eq.
(\ref{eq:WFspinors}) can be related to the relativistic WF already
mentioned in the Introduction. 

Some of the properties discussed in \cite{hinarejos12} can be easily
generalized for the matrix WF. 

1) We have 
\begin{equation}
W_{\beta\alpha}(m,k)=W_{\alpha\beta}^{*}(m,k),
\end{equation}
which implies that the matrix WF is Hermitian. The normalization condition
becomes
\begin{equation}
\sum_{\alpha}\sum_{m}\int_{-\pi}^{+\pi}dkW_{\alpha\alpha}(m,k)=1.
\end{equation}
2) Also, 
\begin{equation}
W_{\alpha\beta}(m,k\pm\pi)=(-1)^{m}W_{\alpha\beta}(m,k).\label{eq:phases}
\end{equation}
3) Given two operators $C$, $D$ and their corresponding Wigner matrices
$W_{\alpha\beta}^{C}(m,k)$, $W_{\alpha\beta}^{D}(m,k)$ one has

\begin{equation}
2\pi\sum_{\alpha,\beta}\sum_{m=-\infty}^{\infty}\int_{-\pi}^{+\pi}dkW_{\alpha\beta}^{C}(m,k)W_{\beta\alpha}^{D}(m,k)=\mathrm{tr}\left(CD\right).\label{eq:innerprod}
\end{equation}
4) A complete knowledge of the WF can be used to reconstruct the density
operator $\rho$:

\begin{equation}
\langle\alpha|\rho|\beta\rangle=2\pi\sum_{m}\int_{-\pi}^{+\pi}dkW_{\alpha\beta}(m,k)A(m,k).\label{eq:rhoWF}
\end{equation}
5) The marginal distributions of (\ref{eq:WFspin}) are related to
matrix elements of the density operator 
\begin{equation}
\sum_{m=-\infty}^{+\infty}W_{\alpha\beta}(m,k)=\tfrac{1}{a}\langle\tfrac{k}{a},\alpha|\rho|\tfrac{k}{a},\beta\rangle,
\end{equation}
and 
\begin{equation}
\int_{-\pi}^{+\pi}dkW_{\alpha\beta}(m,k)=\sum_{n}\delta_{m,2n}\langle n,\alpha|\rho|n,\beta\rangle.
\end{equation}
As already discussed in \cite{hinarejos12}, these equations reflect
the distinction between the coordinates of the phase space points,
$m\in\mathbb{Z}$, $k\in[-\pi,\pi[$, and the position and quasimomentum
bases, $n,$ $q$. The $k$ coordinate is adimensional and does not
directly represent a momentum value, but is connected to $q=k/a$.
The \emph{spatial} label $m$ in phase-space is only connected to
a discrete position, $s$, for even values, $m=2s$, while the odd
values of $m$ are analogous to the odd half-integer phase space grid
points in~\cite{PhysRevA.53.2998,miquel02qc}.

\section{Particular cases}

In order to obtain some insight about the characteristics of the matrix
WF Eq. (\ref{eq:WFspin}), we will give the explicit form it takes
for some particular cases. 
\begin{itemize}
\item Product state
\end{itemize}
We start by considering a product state of spatial and spin degrees
of freedom 
\begin{equation}
\rho=\rho_{L}\otimes\rho_{S},
\end{equation}
where $\rho_{L}$ represents a general state on the lattice, and $\rho_{S}$
is an arbitrary spin state. In this case, we readily obtain
\begin{equation}
W_{\alpha\beta}(m,k)=W_{L}(m,k)\langle\alpha|\rho_{S}|\beta\rangle,\label{eq:WFproduct}
\end{equation}
with 
\begin{equation}
W_{L}(m,k)\equiv\frac{1}{2\pi}\sum_{n}\langle n|\rho_{L}|m-n\rangle e^{-i(2n-m)k}.\label{WFl}
\end{equation}

\begin{itemize}
\item Superposition of two deltas 
\end{itemize}
Let us consider the WM for the state formed by a superposition of
two localized states at lattice sites $\ket{n_{1}}$ and $\ket{n_{2}}$
with $n_{1}\neq n_{2}\in\mathbb{Z}$ 

\begin{equation}
\mid\Psi_{2\delta}\rangle=\frac{1}{\sqrt{1+\mid\alpha\mid^{2}}}(\ket{n_{1}}\mid0\rangle+\alpha\ket{n_{2}}\mid1\rangle),\label{eq:Stat2deltas}
\end{equation}
 where $\alpha$ is an arbitrary complex number that represents the
relative weight of the state $\ket{n_{2}}$. For $\alpha=1$ we obtain
a Schrödinger-cat state. The corresponding WF can be easily calculated.
Written in matrix form in the above spin basis, 

\begin{equation}
W(m,k)=\frac{1}{2\pi(1+\mid\alpha\mid^{2})}\left(\begin{array}{ccc}
\delta_{m,2n_{1}} & \alpha^{*}e^{-ik(n_{1}-n_{2})}\delta_{m,n_{1}+n_{2}}\\
\alpha e^{ik(n_{1}-n_{2})}\delta_{m,n_{1}+n_{2}} & \mid\alpha\mid^{2}\delta_{m,2n_{2}}
\end{array}\right)\label{eq:WF2deltasspin}
\end{equation}
In this case, the WM is zero everywhere except for three particular
values of the space-like phase coordinate, $m=2n_{1},2n_{2},n_{1}+n_{2}$.
It is interesting to compare the structure provided by Eq. (\ref{eq:WF2deltasspin})
with the corresponding superposition of two localized states without
spin \cite{hinarejos12}, given by
\begin{equation}
\mid\Psi_{2\delta}^{no\, spin}\rangle=\frac{1}{\sqrt{1+\mid\alpha\mid^{2}}}(\ket{n_{1}}+\alpha\ket{n_{2}}).
\end{equation}
In that case, the WF is a scalar function 
\begin{eqnarray}
W_{2\delta}^{no\, spin}(m,k) & = & \frac{1}{2\pi(1+|\alpha|^{2})}\left\{ \delta_{m,2n_{1}}+|\alpha|^{2}\delta_{m,2n_{2}}\right.\nonumber \\
 &  & \quad\left.+2|\alpha|\delta_{m,n_{1}+n_{2}}\cos[\Delta n\ k+\phi]\right\} ,\label{eq:WF2deltanospin}
\end{eqnarray}
where $\phi$ is the phase of the complex coefficient $\alpha$, and
$\Delta n=n_{2}-n_{1}$. One observes the different terms in (\ref{eq:WF2deltanospin})
appear distributed on different matrix positions in Eq. (\ref{eq:WF2deltasspin}).
In particular, the out of diagonal term in (\ref{eq:WF2deltasspin})
corresponds to the interference, oscillating term in (\ref{eq:WF2deltanospin}).
This term plays an interesting role related to the non positivity
of the WF. We will return to this point later.
\begin{itemize}
\item Superposition of two Gaussian states 
\end{itemize}
The superposition of two discretized pure Gaussian states with orthogonal
spin components is another interesting state for which the WM defined
in this work can be computed analytically. Such a state is defined
as

\begin{center}
\begin{equation}
|\Psi_{2G}\rangle=\frac{1}{\sqrt{2}{\cal N}}\sum_{n}\left\{ e^{-\frac{(n-a)^{2}}{2\sigma^{2}}}\mid0\rangle+e^{-\frac{(n-b)^{2}}{2\sigma^{2}}}\mid1\rangle\right\} |n\rangle,\label{eq:state2G}
\end{equation}

\par\end{center}

for arbitrary $a,b\in\mathbb{Z}$, $\sigma\in\mathbb{R}^{+}$. For
this state, the WF can be expressed as a matrix in the same $\{\ket 0,\ket 1\}$
basis 

\begin{center}
\begin{equation}
W(m,k)=\frac{1}{2}\left(\begin{array}{ccc}
W_{a}(m,k) & W_{ab}(m,k)\\
W_{ab}^{*}(m,k) & W_{b}(m,k)
\end{array}\right)\label{eq:Wmat2G}
\end{equation}

\par\end{center}

where

\begin{center}
\begin{equation}
W_{l}(m,k)=\frac{1}{2\pi{\cal N}^{2}}e^{-\frac{l^{2}+(m-l)^{2}}{2\sigma^{2}}}e^{ikm}\theta_{3}(k+\frac{im}{2\sigma^{2}},e^{-\frac{1}{\sigma^{2}}}),\,\,\, l=a,b\label{eq:Wa2G}
\end{equation}
 
\par\end{center}

\begin{center}
\begin{equation}
W_{ab}(m,k)=\frac{1}{2\pi{\cal N}^{2}}e^{-\frac{a^{2}+(m-b)^{2}}{2\sigma^{2}}}e^{ikm}\theta_{3}(k+\frac{i(m-b+a)}{2\sigma^{2}},e^{-\frac{1}{\sigma^{2}}})\label{eq:Wab2G}
\end{equation}
 
\par\end{center}

with ${\cal N}=\sqrt{\theta_{3}(0,e^{-\frac{1}{\sigma^{2}}})}$ the
normalization constant. The Jacobi theta function $\theta_{3}(z,q)$
is defined as $\theta_{3}(z,q)\equiv\sum_{n}q^{n^{2}}e^{2izn}$ for
complex arguments $q$, $z$, with $|q|<1$~\cite{abramsteg}. As
in the previous example, we find an important difference with the
WF for the case without spin \cite{hinarejos12}, since the components
in the scalar function appear here distributed as the components of
the matrix WF. In the limit $a=-b\gg\sigma$ with $\sigma\rightarrow0$
we recover the result for the two deltas (\ref{eq:WF2deltasspin})
corresponding to the case $n_{1}=-n_{2}=a$ and $\alpha=1$. 

\begin{figure}
\begin{minipage}[t]{1\columnwidth}%
\includegraphics[width=8cm]{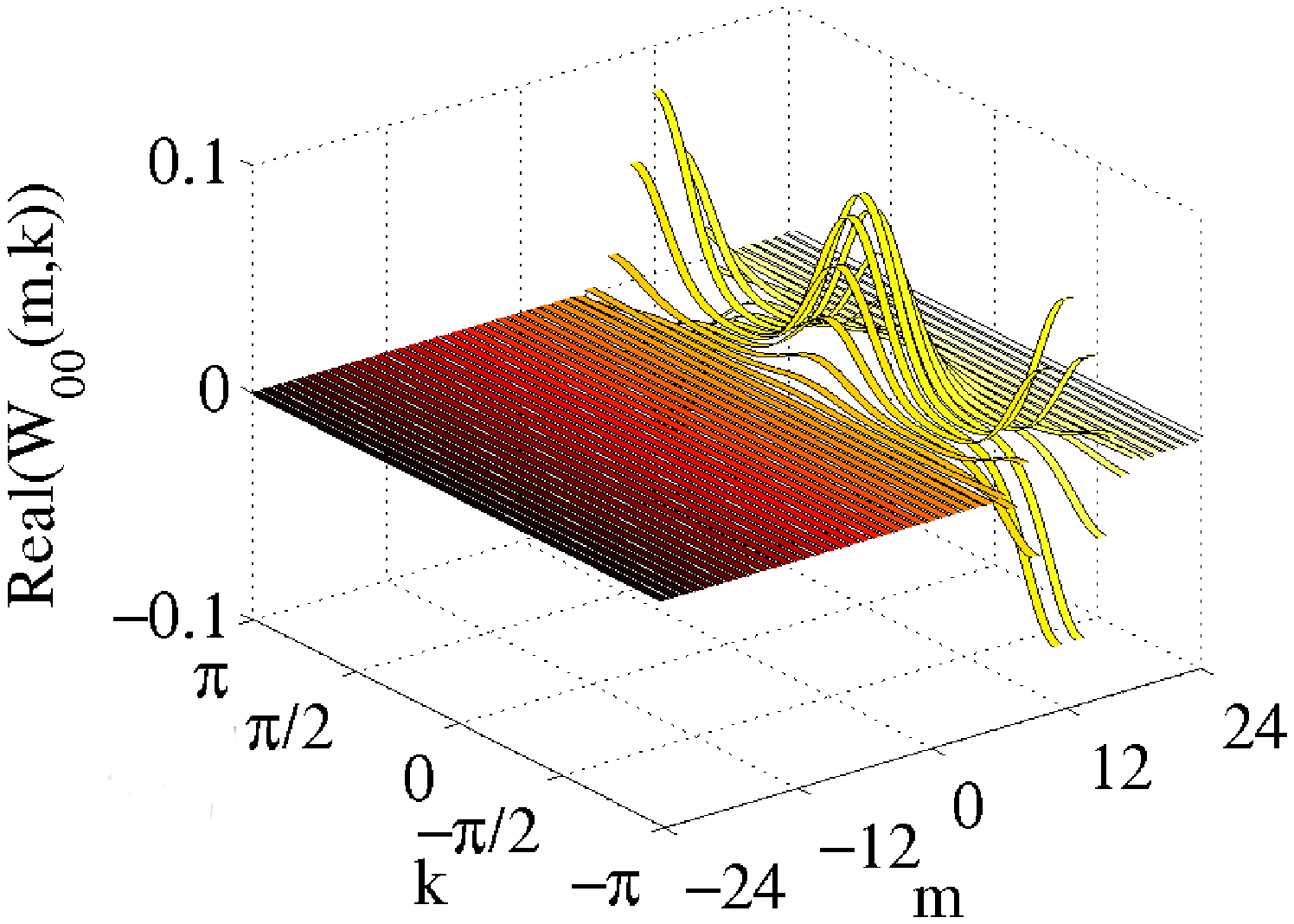}\includegraphics[width=8cm]{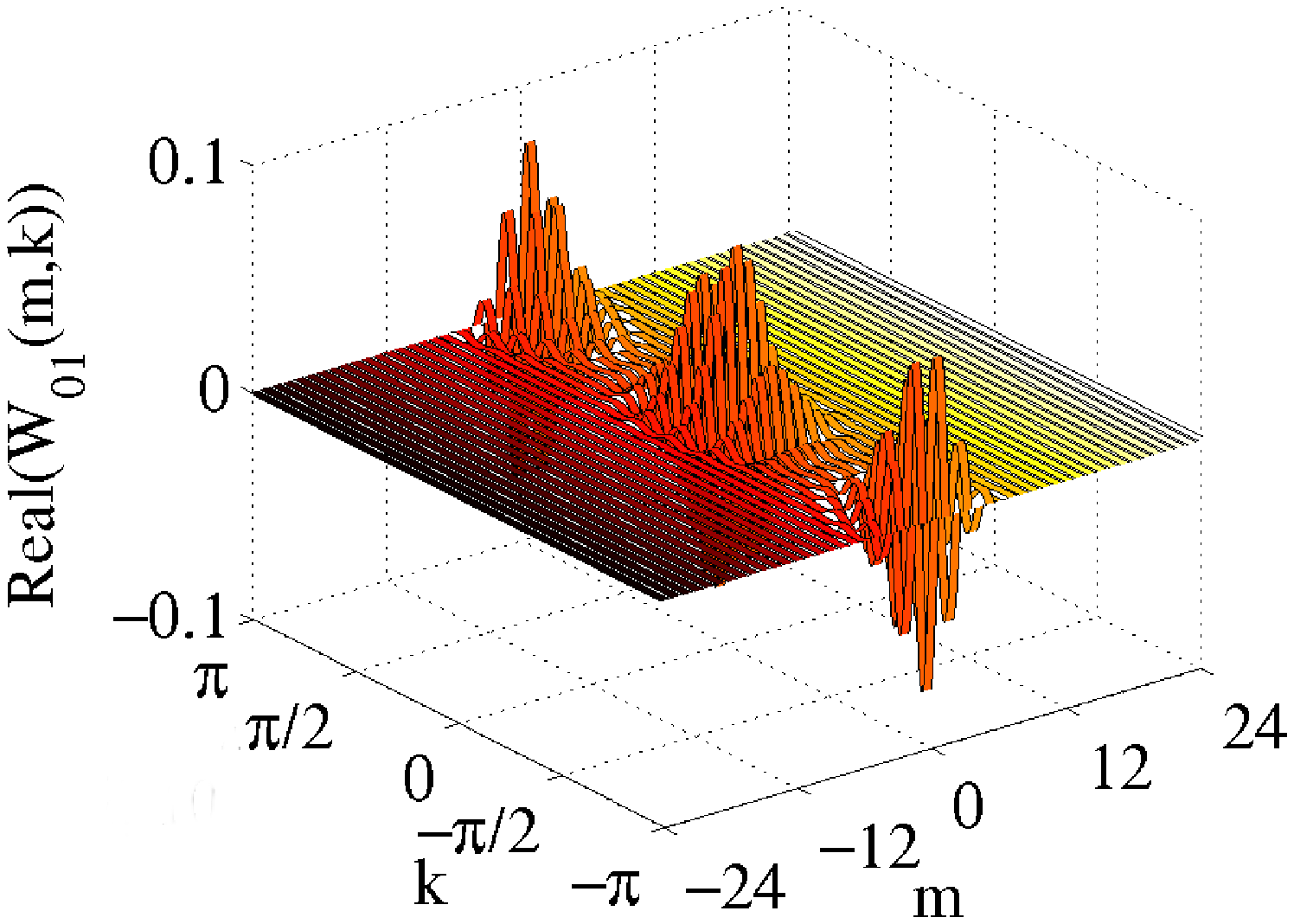}

\includegraphics[width=8cm]{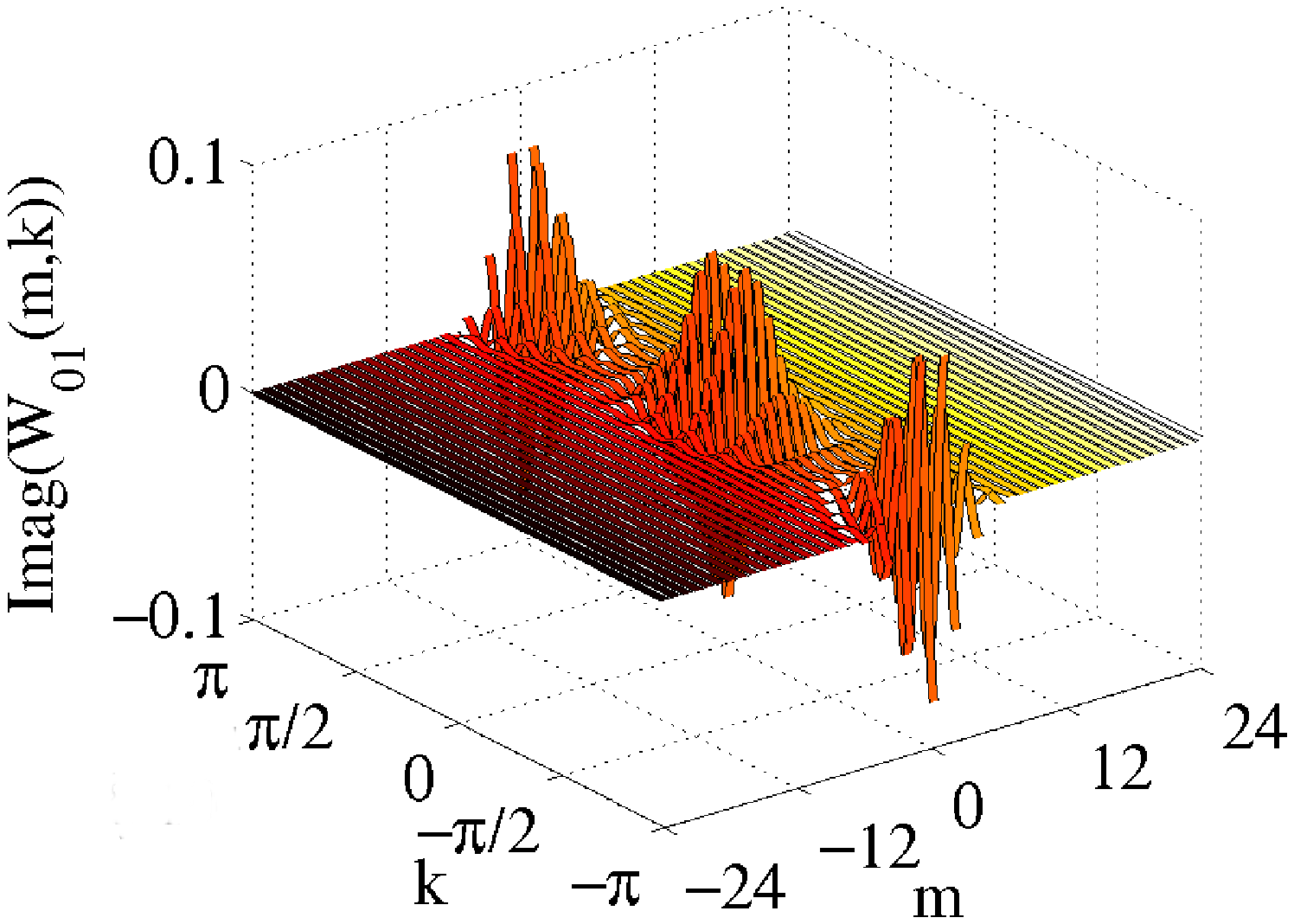}\includegraphics[width=8cm]{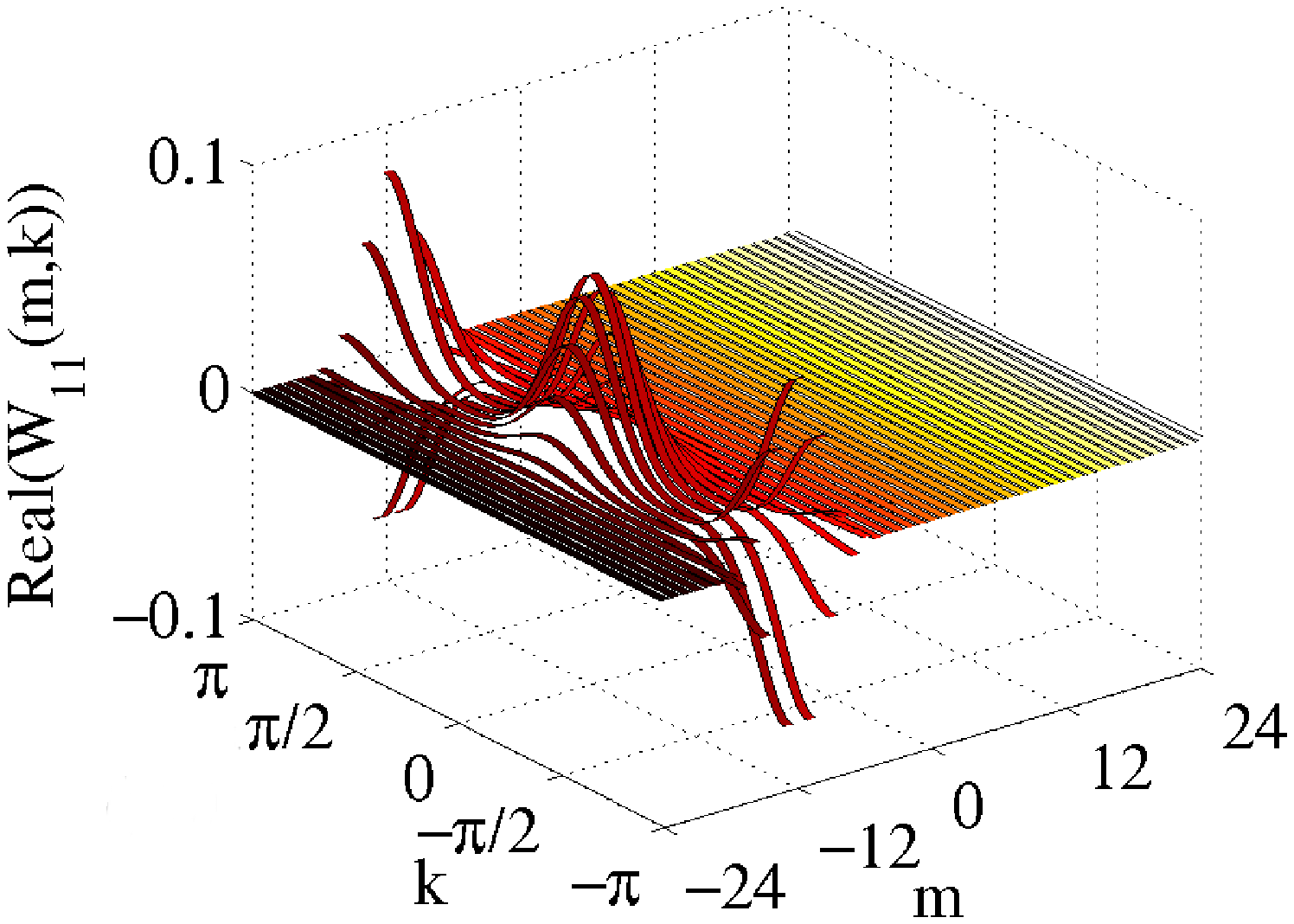}%
\end{minipage}

\caption{Matrix components of the WM for two Gaussians, as given by Eqs. (\ref{eq:Wmat2G}-\ref{eq:Wab2G}).
Panel a) represents \textbf{$W_{a}(m,k)$}, panel b) is the real part
of $W_{ab}(m,k)$, while the imaginary part is plotted on panel c).
Finally, panel d) shows the $W_{b}(m,k)$ component. In these plots,
$a=-b=6$ and $\sigma=1.5$. }

\label{fig2gauss}
\end{figure}

Figure \ref{fig2gauss} shows the four components of the WM for a
two-Gaussian state, as given by Eqs. (\ref{eq:Wmat2G}-\ref{eq:Wab2G}).
One can immediately observe on each component the presence of a secondary
image that reflects the property Eq. (\ref{eq:phases}). In \cite{hinarejos12}
we discussed with some detail, for the spinless case, the peculiarities
related to this duplicate.

\section{Dynamics}

The WF formalism can be used, not only to allow for a description
of a given state, but also to analyze the dynamics, and to visualize
it in phase space. Our purpose is to study the motion of a particle
on a lattice in terms of the corresponding WF. We start from the simplest
case, which corresponds to the spinless particle, and then move to
a more general situation, where the particle interacts with a spin-dependent
term. The time evolution will be first considered within continuous
time, a situation that can be applied to most problems in physics,
and can be described by the Schrödinger equation.

\subsection{Continuous time}

\subsubsection{Particle without spin}

Let us consider a spinless particle moving on a lattice under the
influence of a potential $V$ that depends on the lattice site. We
concentrate on the following Hamiltonian
\begin{equation}
H=J(T_{+}+T_{-})+V,\label{eq:Hlattice}
\end{equation}
that appears as a consequence of the tight-binding approximation in
crystals, where the parameter $J$ is a characteristic of the system
which is related to the hopping probability of an electron to the
nearest neighbor, and the displacement operators $T_{\pm}$ are defined
by $T_{\pm}\ket n=\ket{n\pm1}$. Notice that the Hamiltonian (\ref{eq:Hlattice})
can also be considered as a discretized version of 
\begin{equation}
H_{cont}=-\frac{\nabla^{2}}{2M}+V(x)\label{eq:Hcont}
\end{equation}
(with $M$ the mass of the particle) if one defines $J=-\frac{1}{2Ma^{2}}$. 

The wave function can be written as $\psi(n,t)$, with $t$ the time,
so that the Schrödinger equation %
\footnote{We work in units such that $\hbar=1$%
} reads
\begin{equation}
i\frac{\partial}{\partial t}\psi(n,t)=J\left[\psi(n+1,t)+\psi(n-1,t)-2\psi(n,t)\right]+V_{n}\psi(n,t),\label{eq:Schrolattice}
\end{equation}
with $V_{n}\equiv\braket{n|V|n}$. The last term inside the brackets
in Eq. (\ref{fig2gauss}) can be easily reabsorbed into the definition
of the coefficients $V_{n}$ (it can be also understood as a term
proportional to the identity in the Hamiltonian, thus contributing
only as a position-independent phase as time evolves). Therefore we
omit that term.

It is straightforward to derive an evolution equation satisfied by
the WF for the above problem. We begin with the von Neumann equation
for the density operator
\begin{equation}
\frac{\partial}{\partial t}\rho(t)=-i[H,\rho(t)].\label{eq:von Neumann}
\end{equation}
Making use of (\ref{eq:WFnospin}) one arrives to
\begin{equation}
\frac{\partial}{\partial t}W(m,k,t)=2J\sin k\left[W(m+1,k,t)-W(m-1,k,t)\right]-\frac{i}{2\pi}\sum_{l}e^{-i(2l-m)k}(V_{l}-V_{m-l})\braket{l|\rho(t)|m-l},\label{eq:Wnospintrho}
\end{equation}
where we have explicitly showed the time dependence of $\rho$ and
$W(m,k)$ for the sake of clarity. 

Let us consider that $V(x)$ is a continuous and infinitely derivable
function. In this case, one can obtain a closed form of the above
expression for the WF, as showed in the Appendix. As a result, one
arrives to the following expression

\begin{equation}
\frac{\partial}{\partial t}W(m,k,t)=2J\sin k\left[W(m+1,k,t)-W(m-1,k,t)\right]+\sum_{s=0}^{\infty}\frac{(-1)^{s}a^{2s+1}}{2^{2s}(2s+1)!}\left.\frac{d^{2s+1}V(x)}{dx^{2s+1}}\right|{}_{x=ma/2}\frac{\partial^{2s+1}W(m,k,t)}{\partial k^{2s+1}}.\label{eq:Wnospintser}
\end{equation}
It must be noticed that Eq. (\ref{eq:Wnospintser}) also holds for
the WM (\ref{eq:WFspin}) if we introduce the spin of the particle,
by simply replacing $W(m,k,t)\longrightarrow W_{\alpha\beta}(m,k,t)$,
since none of the spatial operations in this equation can affect the
spin indices. 

Before we go on, we will consider the continuous limit ($a\to0$)
of Eq. (\ref{eq:Wnospintser}). In this limit, our WF has to be replaced
by the corresponding function $W_{\mathrm{c}}(x,q,t)$ following the
prescription \cite{hinarejos12}
\begin{equation}
W(m,k,t)\underset{a\to0}{\longrightarrow}\frac{1}{2}W_{\mathrm{c}}(x=\frac{ma}{2},q=\frac{k}{a},t).\label{eq:Wlimicont}
\end{equation}
By replacing $J=-\frac{1}{2Ma^{2}}$ and substituting (\ref{eq:Wlimicont})
in (\ref{eq:Wnospintser}), and taking the limit ($a\to0$), one obtains
the equation 

\begin{equation}
\frac{\partial}{\partial t}W_{c}(x,q,t)+\frac{q}{M}\frac{\partial}{\partial x}W_{c}(x,q,t)=\sum_{s=0}^{\infty}\frac{(-1)^{s}}{2^{2s}(2s+1)!}\left.\frac{d^{2s+1}V(x)}{dx^{2s+1}}\right|{}_{x=ma/2}\frac{\partial^{2s+1}W_{c}(x,q,t)}{\partial q^{2s+1}}.\label{eq:Wnospintsercont}
\end{equation}
Eq. (\ref{eq:Wnospintsercont}) is the equation of motion for the
WF under the effect of an external potential $V(x)$ in continuous
space, where $q$ represents the momentum of the particle (ranging
from $-\infty$ to $\infty$) (see, for example \cite{Hillery1984121}).

As an interesting particular case, we will study the case of a linear
potential, i.e. $V(x)=\lambda x$, with $\lambda$ a real constant.
Eq. (\ref{eq:Wnospintser}) adopts a simple form
\begin{equation}
\frac{\partial}{\partial t}W(m,k,t)=2J\sin k\left[W(m+1,k,t)-W(m-1,k,t)\right]+\lambda a\frac{\partial}{\partial k}W(m,k,t).\label{eq:Wlinpot}
\end{equation}
To solve this equation, we perform a Fourier transformation on the
variable $m$ by introducing the function
\begin{equation}
\tilde{W}(q,k,t)\equiv\frac{1}{\sqrt{2\pi}}\sum_{m}e^{iqm}W(m,k,t),\label{eq:Wtilde}
\end{equation}
the new variable $q$ taking values on the interval $[-\pi,\pi[$.
With the help of this function, we can rewrite Eq. (\ref{eq:Wtilde})
as 
\begin{equation}
\frac{\partial}{\partial t}\tilde{W}(q,k,t)=-4iJ\sin k\sin q\tilde{W}(q,k,t)+\lambda a\frac{\partial}{\partial k}\tilde{W}(q,k,t).\label{eq:Wtildelin}
\end{equation}
The change of function 
\begin{equation}
\tilde{W}(q,k,t)\equiv e^{-\frac{4iJ\cos k\sin q}{\lambda a}}f(q,k,t)\label{eq:Wtilde factor}
\end{equation}
leads to the following equation for $f(q,k,t)$:

\begin{equation}
\frac{\partial}{\partial t}f(q,k,t)=\lambda a\frac{\partial}{\partial k}f(q,k,t),
\end{equation}
which implies that $f(q,k,t)$ must be of the form $f(q,k,t)=g(q,k+\lambda at),$
with $g(q,k)$ an unknown function that can be determined by the initial
($t=0$) condition in Eq. (\ref{eq:Wtilde factor}), giving
\begin{equation}
g(q,k)=e^{\frac{4iJ\cos k\sin q}{\lambda a}}\tilde{W}(q,k,0).
\end{equation}
We finally obtain, after some algebra
\begin{equation}
\tilde{W}(q,k,t)=\exp[-8i\frac{J}{\lambda a}\sin(k+\frac{\lambda at}{2})\sin(\frac{\lambda at}{2})\sin q]\tilde{W}(q,k+\lambda at,0).\label{eq:Wtildefin}
\end{equation}
To derive an expression for the WF, we need the inverse relation of
Eq. (\ref{eq:Wtilde}), given by

\begin{equation}
W(m,k,t)=\frac{1}{\sqrt{2\pi}}\int_{-\pi}^{\pi}dqe^{-iqm}\tilde{W}(q,k,t),\label{eq:inverseWtilde}
\end{equation}
and make use of the formula \cite{gradshteyn2007} 
\begin{equation}
J_{n}(z)=\frac{1}{2\pi}\int_{-\pi}^{\pi}dqe^{-inq}e^{iz\sin q},\label{eq:Bessel}
\end{equation}
where $n\in\mathbb{Z}$, $z\in\mathbb{C}$, and $J_{n}(z)$ are the
Bessel functions of the first kind. After substituting Eq. (\ref{eq:Wtildefin})
into (\ref{eq:inverseWtilde}) we arrive to the final expression
\begin{equation}
W(m,k,t)=\sum_{l}J_{m-l}\left[-8\frac{J}{\lambda a}\sin(k+\frac{\lambda at}{2})\sin\frac{\lambda at}{2}\right]W(l,k+\lambda at,0).\label{eq:Wexlinpot}
\end{equation}
Notice that, in the latter equation, the argument $k+\lambda at$
is to be understood modulo $2\pi$. Using this fact, one can readily
obtain that the above solution exhibits a time periodicity
\begin{equation}
W(m,k,t+\frac{2\pi}{\lambda a})=W(m,k,t),
\end{equation}
which corresponds to the well known phenomenon of Bloch oscillations,
that can be observed for electrons confined in a periodic potential
(the lattice) subject to a constant force, as for example a constant
electric field. The corresponding frequency $\omega_{B}=\abs{\lambda}a$
is precisely what is expected for our linear potential $V(x)=\lambda x$.

Directly related to the above treatment, it appears quite natural
to attempt a parallelism with a situation that describes the dynamics
of a particle under the effect of a constant gravitational field,
$V(x)=m_{g}gx$, where $m_{g}$ is the gravitational mass and $g$
the acceleration of gravity. Notice that, for the following discussion
to make sense, one should design a physical system that is described
by this potential, and that Eq. (\ref{eq:Schrolattice}) can be considered
as a discretized approximation to (\ref{eq:Hcont}), with $J=-\frac{1}{2Ma^{2}}$
. We will return to this discussion later.

We find it convenient to use the symbol $m_{i}$ instead of $M$ to
represent the \textit{inertial mass,} and to recover the Planck constant.
We observe that the argument of the Bessel functions in Eq. (\ref{eq:Wexlinpot})
depends upon the combination

\begin{equation}
-\frac{J}{\lambda a}=\frac{\hbar^{2}}{2m_{i}m_{g}ga^{3}}\equiv\frac{1}{(k_{g}a)^{3}},
\end{equation}
where $k_{g}\equiv(\frac{2m_{i}m_{g}g}{\hbar^{2}})^{1/3}$ is a characteristic
wave vector that modulates the spatial dependence of energy eigenstates
in a gravitational field in continuous space \cite{kajari2010}. As
the authors of this work discuss, this is one of the possible effects
for quantum particles under the effect of gravity, where various combinations
of (powers of) $m_{g}$ and $m_{i}$ may appear depending on the problem
under consideration, thus paving the way to measuring these two quantities
independently. 

The dynamics on the lattice we just considered offers a similar perspective.
The time evolution in Eq. (\ref{eq:Wexlinpot}) is governed by the
product $k_{g}a$, which involves the lattice spacing as a new parameter,
thus allowing an extra degree of freedom in the design of experiments,
if they are performed on a lattice instead of in continuous space.
However, one has to be careful about this point: Only if the design
of the experiment is such that $J$ and $V(x)$ correspond to the
above hypothesis, the previous discussion can make sense. 

\begin{figure}
\includegraphics[width=9cm]{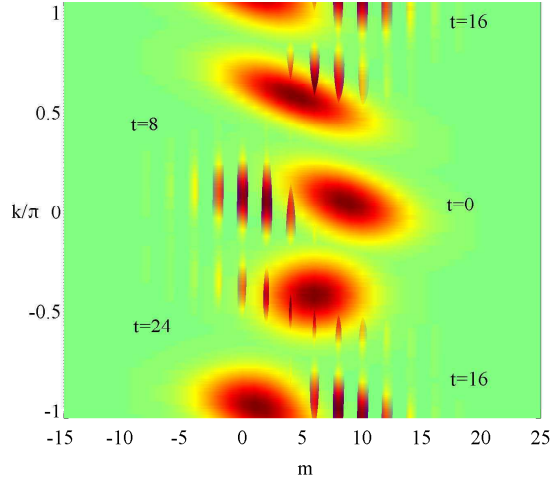}

\caption{Snapshots corresponding to the time evolution of the WF, as given
by Eq. (\ref{eq:Wexlinpot}) with an initial Gaussian state of the
form (\ref{eq:Wa2G}) with $a=3$ and $\sigma=2$. The parameters
of the Hamiltonian are $J=1$ and $\lambda a=1$. The labels indicate
different values of time.}

\label{figWFtimesonspin}
\end{figure}

To illustrate the behavior of the WF, we plotted in Fig. \ref{figWFtimesonspin}
several snapshots obtained by evolving an initial Gaussian state of
the form (\ref{eq:Wa2G}). The time evolution is governed by Eq. (\ref{eq:Wexlinpot}).
One observes several features on this plot. First, the position of
the maximum shows oscillations for the variable $m$, as corresponding
to the Bloch oscillations discussed above, while variable $k$ evolves
linearly (and periodically) with time. During the evolution, the WF
also experiences a distortion that is similar to the one observed
in continuous space \cite{kajari2010}. One also observes the presence
of a secondary image which manifests as vertical strips.

\subsubsection{Particle with spin}

We return to the description of a particle with spin 1/2. Our purpose
is to analyze the dynamics for such a system, and compare it with
the spinless case. To do so, we need to introduce some spin-dependent
potential, otherwise the different components in the WM will evolve
exactly in the same way, and the results of the previous subsection
apply. In order to make this comparison as close as possible, we will
consider the time evolution under the effect of a Hamiltonian of the
form

\begin{equation}
H=J(T_{+}+T_{-})+\sigma_{z}V,\label{eq:Hlatticespin}
\end{equation}
where $V$ is, as before, a site-dependent scalar potential. It is
possible to obtain an evolution equation, similar to (\ref{eq:Wnospintser}),
when the particle is subject to the above Hamiltonian in the lattice.
This derivation is made in the Appendix, the main difference with
the spinless case being that the diagonal and off-diagonal components
of the WM evolve differently. In what follows, we concentrate on the
particular example of a discretized linear potential $V_{n}=\lambda an$,
with $\lambda$ a real constant. Then, Eq. (\ref{eq:WnospintrhoDspin})
particularizes to 
\begin{equation}
\frac{\partial}{\partial t}W_{\alpha\alpha}(m,k,t)=2J\sin k\left[W_{\alpha\alpha}(m+1,k,t)-W_{\alpha\alpha}(m-1,k,t)\right]+(-1)^{\alpha}\lambda a\frac{\partial}{\partial k}W_{\alpha\alpha}(m,k,t),\label{eq:Wlinpotdiag}
\end{equation}
and

\begin{equation}
\frac{\partial}{\partial t}W_{\alpha\beta}(m,k,t)=2J\sin k\left[W_{\alpha\beta}(m+1,k,t)-W_{\alpha\beta}(m-1,k,t)\right]-i(-1)^{\alpha}\lambda amW_{\alpha\beta}(m,k,t),\label{eq:Wlinpotnodiag}
\end{equation}
(valid for $\alpha\neq\beta$).

The first equation can be easily solved by comparison to (\ref{eq:Wlinpot}).
We only have to perform the replacement $\lambda\longrightarrow(-1)^{\alpha}\lambda$.
Therefore, we can write the solution using the same procedure as in
the case with no spin, to obtain
\begin{equation}
W_{\alpha\alpha}(m,k,t)=\sum_{l}J_{m-l}\left[-8\frac{J}{\lambda a}\sin(k+(-1)^{\alpha}\frac{\lambda at}{2})\sin\frac{\lambda at}{2}\right]W_{\alpha\alpha}(l,k+(-1)^{\alpha}\lambda at,0).\label{eq:Wexlinpotspindiag}
\end{equation}
The same comments made in the previous section hold here: $W_{\alpha\alpha}(m,k,t)$
is periodic in time, with frequency given by $\omega_{B}=\abs{\lambda}a$
. Eq. (\ref{eq:Wlinpotnodiag}) can be solved by introducing a Fourier
transform, as made with (\ref{eq:Wlinpot}). We arrive, after some
algebra, at 
\begin{equation}
W_{\alpha\beta}(m,k,t)=e^{(-1)^{\alpha}im\frac{\lambda at}{2}}\sum_{l}e^{(-1)^{\alpha}il\frac{\lambda at}{2}}J_{m-l}\left[-8\frac{J}{\lambda a}\sin k\sin\frac{\lambda at}{2}\right]W_{\alpha\beta}(l,k,0).\label{eq:Wexlinpot-1}
\end{equation}
(valid when $\alpha\neq\beta$).

\begin{figure}
\begin{minipage}[t]{1\columnwidth}%
\includegraphics[width=9cm]{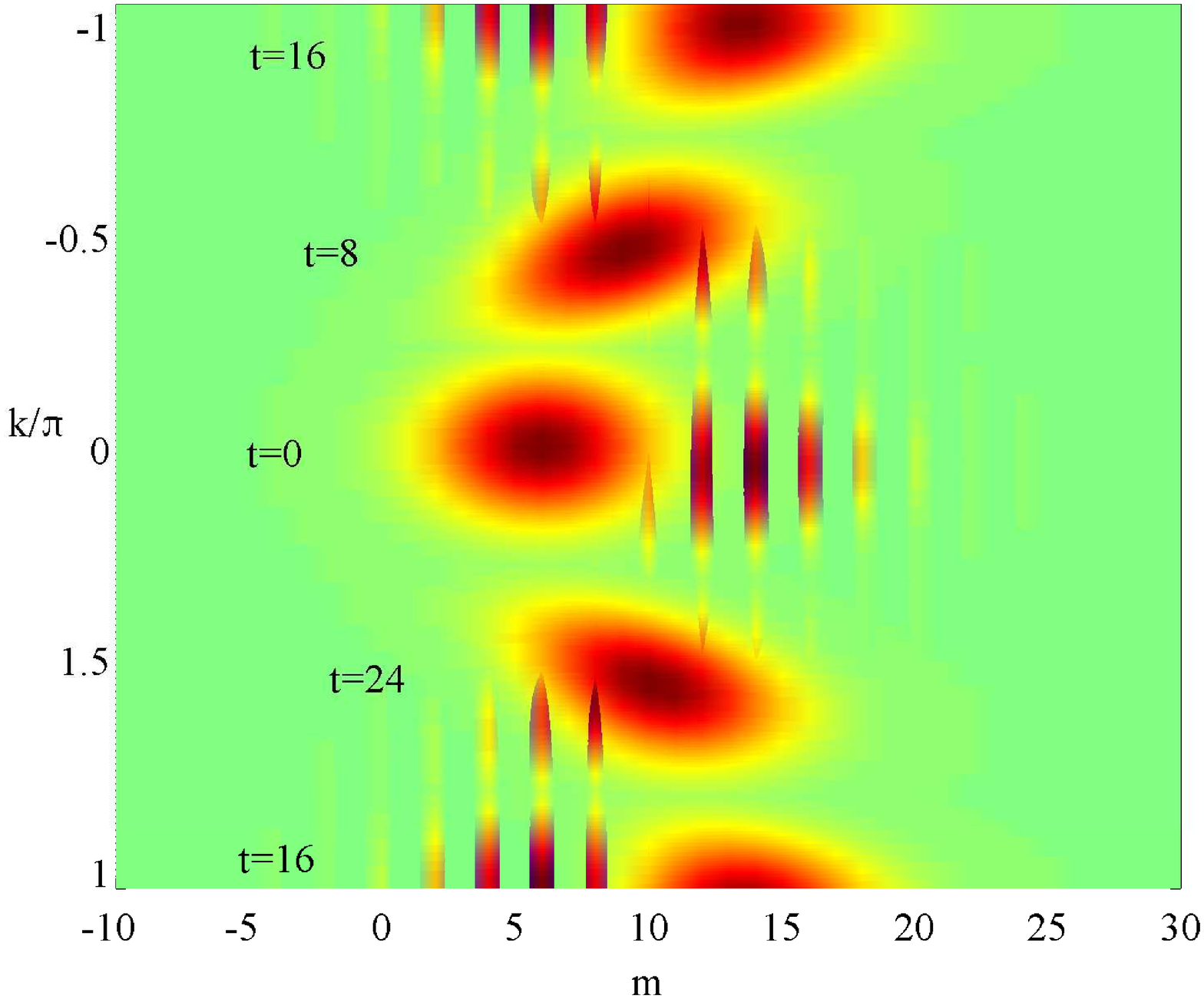}\includegraphics[width=9cm]{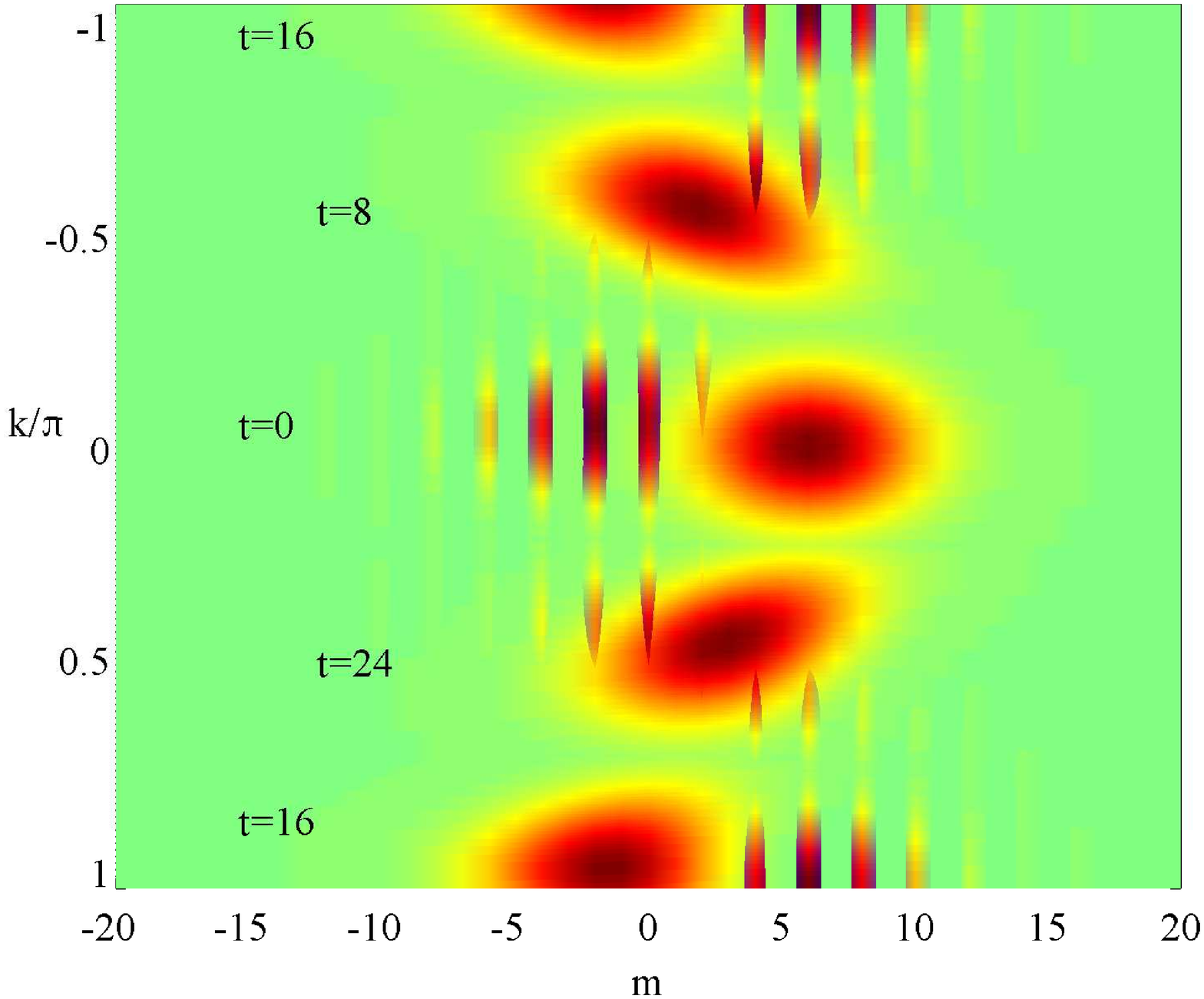}%
\end{minipage}

\caption{The two panels show the diagonal components of the WM at four different
times (labeled by the corresponding $t$), for a particle subject
to the interaction Hamiltonian (\ref{eq:Hlatticespin}) in a lattice.
Left panel corresponds to $W_{00}(m,k,t)$, whereas right panel shows
$W_{11}(m,k,t)$. The initial state is a separable state (see the
main text for explanation) with $a=3$, $\sigma=2$. The parameters
of the interaction Hamiltonian are $J=\lambda a=1$.}

\label{figGaussspin}
\end{figure}

To illustrate the evolution of the WM elements under the effect of
the Hamiltonian (\ref{eq:Hlatticespin}) with a linear potential,
we followed this evolution for an initial separable state of the form
(\ref{eq:WFproduct}), with $\rho_{S}$ defined by the pure state
$\frac{1}{\sqrt{2}}(\ket 0+\ket 1)$ and $W_{L}(m,k)$ corresponding
to a Gaussian state, given by (c.f. Eq. (\ref{eq:Wa2G}))
\begin{equation}
W_{L}(m,k)=\frac{1}{2\pi{\cal N}^{2}}e^{-\frac{a^{2}+(m-a)^{2}}{2\sigma^{2}}}e^{ikm}\theta_{3}(k+\frac{im}{2\sigma^{2}},e^{-\frac{1}{\sigma^{2}}})\label{eq:WLG}
\end{equation}
and ${\cal N}=\sqrt{\theta_{3}(0,e^{-\frac{1}{\sigma^{2}})}}$ the
normalization constant. The results are presented in Fig. \ref{figGaussspin},
which shows different snapshots of the diagonal components $W_{00}(m,k,t)$
and $W_{11}(m,k,t)$ of the WM. We observe that both components present
similar features to the case without spin, plotted in Fig. \ref{figWFtimesonspin}.
However, they evolve differently on the $m$ axis: Initially, the
$W_{00}(m,k,t)$ component moves to the left, while the $W_{11}(m,k,t)$
component moves to the right, as a consequence of the different time
dependence $(-1)^{\alpha}\lambda at$ in (\ref{eq:Wexlinpotspindiag}),
a phenomenon which is reminiscent of the splitting into two beams
on the Stern-Gerlach experiment, where the basic piece of the interaction
is analogous to (\ref{eq:Hlatticespin}).

\subsubsection{Decoherence}

Another dynamical scenario of great relevance for the study of quantum
systems is the presence of decoherence, which can be caused by interaction
with the environment. In the following we show how the WF formalism
we are discussing accommodates also such situation. In particular,
we explore some typical cases, in which the spin structure of the
WM allows a simple visualization of the decoherence effects.

We consider the case where the interaction with the environment can
be described by a Lindblad-type equation \cite{Breuer2007}
\begin{equation}
\frac{\partial}{\partial t}\rho=-i[H,\rho]+\sum_{k}\gamma_{k}(A_{k}\rho A_{k}^{\dagger}-\frac{1}{2}A_{k}^{\dagger}A_{k}\rho-\frac{1}{2}\rho A_{k}^{\dagger}A_{k}),\label{eq:Lindblad}
\end{equation}
where $A_{k}$ are the Lindblad operators, and $\gamma_{k}$ represent
the corresponding coupling constants. 

If these operators act only on the spin space, the Lindblad (noise)
term Eq. (\ref{eq:Lindblad}) immediate translates in an analogous
equation for the WM. In other words, under this hypothesis we can
write for the WM 
\begin{equation}
\frac{\partial}{\partial t}W(m,k,t)=\frac{\partial}{\partial t}W(m,k,t)\vert_{H}+\sum_{k}\gamma_{k}(A_{k}W(m,k,t)A_{k}^{\dagger}-\frac{1}{2}A_{k}^{\dagger}A_{k}W(m,k,t)-\frac{1}{2}W(m,k,t)A_{k}^{\dagger}A_{k}).\label{eq:LindbladWF}
\end{equation}
In the latter equation, $\frac{\partial}{\partial t}W(m,k,t)\vert_{H}$
denotes the contribution of the Hamiltonian to the dynamics (without
decoherence), and we used a matrix notation, so that spin indices
are omitted.

As a simple example, let us consider the case when we only have a
Lindblad operator $A_{1}=\sigma_{z}$ with $\gamma_{1}\equiv\gamma$.
We then have 

\begin{equation}
\frac{\partial}{\partial t}W(m,k,t)=\frac{\partial}{\partial t}W(m,k,t)\vert_{H}+\left(\begin{array}{cc}
0 & -2\gamma W_{01}(m,k,t)\\
-2\gamma W_{10}(m,k,t) & 0
\end{array}\right),\label{eq:LindbladWFsz}
\end{equation}
which solution can be readily obtained, and expressed as
\begin{equation}
W(m,k,t)=\left(\begin{array}{cc}
W_{00}(m,k,t)\vert_{H} & e^{-2\gamma t}W_{01}(m,k,t)\vert_{H}\\
e^{-2\gamma t}W_{10}(m,k,t)\vert_{H} & W_{11}(m,k,t)\vert_{H}
\end{array}\right).
\end{equation}
In other words, in this example decoherence leaves the diagonal terms
unaltered, while the off-diagonal terms are exponentially damped with
time.

Our second example is provided by the Lindblad operator $A_{1}=\sigma_{x}$
with $\gamma_{1}\equiv\gamma$. In this case, Eq. (\ref{eq:LindbladWF})
becomes

\begin{equation}
\frac{\partial}{\partial t}W(m,k,t)=\frac{\partial}{\partial t}W(m,k,t)\vert_{H}+\gamma\left(\begin{array}{cc}
W_{11}(m,k,t)-W_{00}(m,k,t) & W_{10}(m,k,t)-W_{01}(m,k,t)\\
W_{01}(m,k,t)-W_{10}(m,k,t) & W_{00}(m,k,t)-W_{11}(m,k,t)
\end{array}\right).\label{eq:LindbladWFsx}
\end{equation}
This set of equations can be solved by elementary operations. We concentrate
on the diagonal terms, for which the final solution reads
\begin{equation}
W_{00}(m,k,t)=\frac{1}{2}(1+e^{-2\gamma t})W_{00}(m,k,t)\vert_{H}+\frac{1}{2}(1-e^{-2\gamma t})W_{11}(m,k,t)\vert_{H},
\end{equation}

\begin{equation}
W_{11}(m,k,t)=\frac{1}{2}(1-e^{-2\gamma t})W_{00}(m,k,t)\vert_{H}+\frac{1}{2}(1+e^{-2\gamma t})W_{11}(m,k,t)\vert_{H}.
\end{equation}
Similar equations can be obtained involving $W_{01}(m,k,t)$ and $W_{10}(m,k,t)$.
As a result, in the limit $t\longrightarrow\infty$ both $W_{00}(m,k,t)$
and $W_{11}(m,k,t)$ become an equally weighted mixture (the same
happens with the off-diagonal terms).

\subsection{Discrete time}

\subsubsection{Quantum walk}

The examples studied in the previous Section arise as a consequence
of the continuous interaction of a particle with an external potential
acting on the lattice. However, we can envisage some situations in
which we act on the particle with subsequent short pulses, or via
some actions that appear suddenly, but regularly in time. A paradigmatic
example of this kind is provided by the quantum walk \cite{Aharonov93,Venegas-Andrac2012},
which has received a lot of interest in recent years. In the discrete
quantum walk, a quantum particle moves on an (1D) lattice subject
to the periodic influence of a displacement operator, that propagates
the particle to the right or to the left, according to the state of
a two-level system (the \textit{coin}). The total Hilbert space has
precisely the structure $\mathcal{H}=\mathcal{H}_{l}\otimes\mathcal{H}_{s}$,
defined in Sect. II and, in fact, we can associate the states of the
coin to the spin of the particle, without loss of generality. It is
customary to use the basis states $\ket L$ and $\ket R$ in $\mathcal{H}_{s}$
(instead of $\ket 0$ and $\ket 1$) and associate them to the left
and right propagation, respectively. We consider the successive application
of the unitary transformation 
\begin{equation}
U(\theta)=\left\{ T_{-}\otimes|L\rangle\langle L|+T_{+}\otimes|R\rangle\langle R|\right\} \otimes C(\theta),\label{Ugen}
\end{equation}
where $C(\theta)=I\otimes\sigma_{z}e^{-i\theta\sigma_{y}}$, $\theta\in\left[0,\pi/2\right]$
is a parameter defining the bias of the coin toss, $I$ is the identity
operator in $\mathcal{H}_{l}$, and $\sigma_{y}$ and $\sigma_{z}$
are Pauli matrices acting on $\mathcal{H}_{s}$. The QW dynamics can
be described entirely in terms of the WM \cite{Hinarejos2013}, via
a recursion formula that relates $W(m,k,t+1)$ to other components
of this function at time $t$. Using Eq. (\ref{Ugen}) one obtains,
after some algebra:

\selectlanguage{spanish}%
\begin{align}
W(m,k,t+1) & =M_{R}W(m-2,k,t)M_{R}^{\dagger}+e^{-2ik}M_{R}W(m,k,t)M_{L}^{\dagger}\nonumber \\
+ & e^{2ik}M_{L}W(m,k,t)M_{R}^{\dagger}+M_{L}W(m+2,k,t)M_{L}^{\dagger},\label{recWigner}
\end{align}
\foreignlanguage{english}{where $M_{L}=(|L\rangle\langle L|)C(\theta)$
and $M_{L}=(|R\rangle\langle R|)C(\theta)$. A complete analysis of
the time evolution in phase space with the help of the WF can be found
in \cite{Hinarejos2013}. Notice that a different definition of the
WF was used in \cite{Lopez2003} for the reduced density matrix of
the walker (after tracing the coin) to study of the evolution and
the effects of decoherence for the quantum walk.}

\selectlanguage{english}%

\subsubsection{Decoherence in discrete time }

The WF formalism can easily accommodate the description of the general
transformation of the quantum state via a completely positive (CP)
map. In particular, we consider here trace preserving maps. These
could, for instance, represent a decoherent QW process, with Kraus
operators modeling the interaction of the system with the environment.
The discrete evolution is represented by
\begin{equation}
\rho(t+1)=\sum_{i}E_{i}\rho(t)E_{i}^{\dagger},\label{eq:mapE}
\end{equation}
where $E_{i}$ are Kraus operators with the property $\sum_{i}E_{i}^{\dagger}E_{i}=I$.
As an example, we analyze two simple models of decoherence which are
applied as projective measurements in the different degrees of freedom
of the system. The first model is defined as projectors in spin space,
while the second model is defined by projecting in the lattice sites.
We use the notation $\Pi_{i}$ to designate the different projectors,
which satisfy $\Pi_{i}^{\dagger}=\Pi_{i}$ and $\Pi_{i}\Pi_{j}=\delta_{ij}\Pi_{i}$.
With probability $p$, the system is projected onto the spin (or space)
basis, so that Eq. (\ref{eq:mapE}) will be rewritten as
\begin{equation}
\rho(t+1)=(1-p)\rho(t)+p\sum_{i}\Pi_{i}\rho(t)\Pi_{i}.\label{eq:mapPi}
\end{equation}
By iteration of the above equation and making use of the properties
of projectors, one can derive the following formula relating the final
and initial density operators of the system,

\begin{equation}
\rho(t)=(1-p)^{t}\rho(0)+[1-(1-p)^{t}]\sum_{i}\Pi_{i}\rho(t)\Pi_{i}.\label{eq:rhoaftert}
\end{equation}
We start from a state consisting of superposition of two deltas with
orthogonal spin components, Eq. (\ref{eq:Stat2deltas}) with $\alpha=1$.
For the first projective model we apply the spin projectors $\Pi_{i}=\ket i\bra i$,
$i=0,1$, while for the site projection they are given by $\Pi_{n}=\ket n\bra n,n\in\mathbb{Z}$.
The iterated density operator $\rho(t)$ that is obtained from Eq.
(\ref{eq:rhoaftert}) is the same in both cases, the reason being
the spin and position entanglement structure in Eq. (\ref{eq:Stat2deltas}).
The result is

\begin{equation}
\rho(t)=\frac{1}{2}\left(\begin{array}{ccc}
\mid n_{1}\rangle\langle n_{1}\mid & (1-p)^{t}\mid n_{1}\rangle\langle n_{2}\mid\\
(1-p)^{t}\mid n_{2}\rangle\langle n_{1}\mid & \mid n_{2}\rangle\langle n_{2}\mid
\end{array}\right).
\end{equation}
The corresponding WM becomes 

\begin{equation}
W(m,k,t)=\frac{1}{4\pi}\left(\begin{array}{ccc}
\delta_{m,2n_{1}} & (1-p)^{t}\delta_{m,n_{1}+n_{2}}e^{-ik(n1-n2)}\\
(1-p)^{t}\delta_{m,n_{1}+n_{2}}e^{ik(n1-n2)} & \delta_{m,2n_{2}}
\end{array}\right).\label{eq:WFiterated}
\end{equation}
Thus, as a consequence of the projective measurements, the non-diagonal
components in the WM (\ref{eq:WFiterated}) tend to zero with time.
This was expected from the intuitive idea that these components appear
from interference between the two spin states in Eq. (\ref{eq:Stat2deltas})
(or, correspondingly, between the two occupied positions): Once decoherence
acts, this kind of interference is reduced and the responsible terms
are consequently diminished. Qualitatively similar results are found
if one starts from the superposition of two Gaussian states (\ref{eq:state2G}),
and introduces projective measurements on the lattice states. Interestingly,
these interference terms are non positive and tend to disappear as
decoherence is acting. We will discuss the consequences of this idea
in the next Section with more detail.

\section{Negativity}

In the context of continuous variables, it is well known that the
Wigner function may present some zones in phase space where it is
negative. This is interpreted as an indication of quantumness, in
the sense that the state would not have a classical analogue. In order
to quantify this quantum feature, the negative volume of the Wigner
function has been defined as a measure on non-classicality \cite{kenfack}
and has been applied to distinguish quantum states from classical
ones \cite{PhysRevLett.106.010403}. The only pure states with non-negative
Wigner function are Gaussian states \cite{hudson74}, however the
classification is not complete for mixed states.

For the continuous phase space, the negativity of a state $\rho$
becomes 
\begin{equation}
\eta(\rho)=\int_{-\infty}^{\infty}\int_{-\infty}^{\infty}[\mid W_{c}(x,p)\mid-W_{c}(x,p)]dpdx=\int_{-\infty}^{\infty}\int_{-\infty}^{\infty}\mid W_{c}(x,p)\mid dpdx-1.\label{negscalar}
\end{equation}

The positive character of the Wigner function has also been studied
for discrete systems. In the finite dimensional case, and for odd
dimension, Gross showed \cite{gross06hudson} that the only pure states
with positive Wigner function are stabilizer states. The presence
of negative values in the Wigner function has been in this case connected
to a quantum resource, related to a possible quantum speedup \cite{cormick06class,galvao05speedup}
or the non-simulability of certain quantum computations involving
states with non-positive Wigner function \cite{mari12,veitch12}.

In the case of spin $\frac{1}{2}$, the Wigner function defined by
Wooters \cite{Wootters1987} has been used to establish a separability
criterion for a system of two particles \cite{Franco2006}. A connection
between entanglement and negative Wigner functions was established
also in \cite{PhysRevLett.106.010403} for two particles in a continuous
space, when the state is a hyperradial s-wave.

Even without the additional degree of freedom, the discreteness of
the Hilbert space causes the appearance of spurious negative terms
in the Wigner function, which do not correspond directly to non-classical
features of the state, but are due to the structure of the discrete
phase space itself. Nevertheless, for the case of a spinless particle
we showed in \cite{hinarejos12} that it is possible to introduce
a modified negativity measure which excludes such negative contributions
and contains information about the quantumness of the states, consistent
with the continuum limit. 

Not being a true quantum observable, the meaning of such a negativity
measure will depend strongly on the definition used for the Wigner
function and on the characteristics of the particular system, as the
discussion above illustrates. It is then reasonable to ask what natural
extension corresponds to the system we are discussing, and what information
it maintains about the characteristics of the states. 

It is possible to think of several extensions of the Wigner function.
If we start with our definition (\ref{eq:WFspin}) and trace out the
spin, we are left with a scalar Wigner function representing the state
of the spatial degree of freedom, which in general will be mixed.
To this function we can immediately apply the definition of negativity
discussed in \cite{hinarejos12}. It might be more interesting to
think of a negativity definition $\eta(\rho)$ that preserves some
spin information.

One possibility is to define a negativity for the Wigner matrix, as
in \cite{Hinarejos2013}, 
\begin{equation}
\eta(\rho)\equiv\sum_{m}\int_{-\pi}^{\pi}[\mid\mid W(m,k)\mid\mid_{1}-Tr(W(m,k))]dk=\sum_{m}\int_{-\pi}^{\pi}\mid\mid W(m,k)\mid\mid_{1}dk-1,\label{eq:negativity}
\end{equation}
where $\mid\mid A\mid\mid_{1}\equiv Tr\sqrt{A^{\dagger}A}$ is the
trace norm of matrix $A$, and the second equality follows from normalization.
We can easily check that this quantity fulfills the following desirable
properties: 
\begin{enumerate}
\item It reduces to Eq. (\ref{negscalar}) for product states in the continuum
limit, with $W_{c}(x,p)$ obtained from $W_{L}(m,k)$ (see Eq. (\ref{eq:WFproduct}); 
\item It is invariant under rotations in spin space. 
\end{enumerate}
The first property is also satisfied by the negativity computed after
tracing out the spin. The second property, on the other hand, can
be illustrated with the following example. We consider an electron,
subject to an external magnetic field. To simplify, the electron is
confined to a site on the lattice, so that its state is factorizable.
The effect of the magnetic field manifests on the precession of the
spin, which continuously changes the spin state of the electron. This
property ensures that the value of the negativity is not influenced
by the precession. In other words, simply changing the spin direction
will not alter the negativity properties of the Wigner matrix. Notice
that, for some alternative definitions of the Wigner function for
a particle with spin \cite{Wootters1987}, the function can contain
negative values in the phase space for some states, while being completely
positive for other states.

We can further explore the significance of the definition (\ref{eq:negativity})
by considering different examples. We may then investigate, as in
\cite{Franco2006}, whether this quantity holds information about
the entanglement in the state.

We start by analyzing the cat state, $\mid\psi\rangle=\frac{1}{\sqrt{1+\mid\beta\mid^{2}}}(\mid a\rangle\ket{\sigma_{1}}+\beta\mid b\rangle\mid\sigma_{2}\rangle)$
where $a,b\in\mathcal{Z}$ label two different sites on the lattice,
$\beta\in\mathbb{C}$ is a constant, and $\{\ket{\sigma_{1}},\ket{\sigma_{2}}\}$
are two arbitrary, orthogonal spin states. The negativity of this
state takes the form: $\eta=\frac{2\mid\beta\mid}{1+\mid\beta\mid^{2}}$.
It is easy to check that in this case the entanglement and the negativity
have the same behavior.

However, this is not the generic behavior, as illustrated by Werner
states \cite{PhysRevA.40.4277}, $\rho=\frac{1-z}{4}I+z\mid\psi\rangle\langle\psi\mid$,
where $\mid\psi\rangle=\frac{1}{\sqrt{2}}(\mid a\rangle\ket 0+\mid b\rangle\ket 1)$
and $a,b\in\mathcal{Z}$ label two different sites on the lattice.
This state is entangled whenever $z\geq\frac{1}{3}$. The Wigner matrix
for this state takes the form 
\begin{equation}
W(m,k)=\left(\begin{array}{cc}
\frac{1+z}{4}W_{aa}(m,k)+\frac{1-z}{4}W_{bb}(m,k) & \frac{z}{2}W_{ab}(m,k)\\
\frac{z}{2}W_{ab}(m,k) & \frac{1-z}{4}W_{aa}(m,k)+\frac{1+z}{4}W_{bb}(m,k)
\end{array}\right),
\end{equation}
with the definition $W_{ln}(m,k)=\frac{1}{2\pi}\delta_{m,l+n}e^{-ik(l-n)}$
and $l,n\in\{a,b\}$. The corresponding negativity is simply $\eta(\rho)=z$.
This result implies that for these states, entanglement and negativity
are not correlated.

Another scenario where the emergence of classical behavior is often
discussed is that of decoherent dynamics. It is thus reasonable to
study how this quantity, $\eta$, changes under decoherence. To do
so, we consider a very simple situation, in which the initial state
subject to decoherence is the double delta considered in Sect. III.
To simplify, we restrict ourselves to the discrete time dynamics already
studied in Sect. IV, with decoherence arising from projections on
spin or lattice sites. Similar qualitative conclusions can be drawn
if we allow for a continuous time dynamics, or if we consider a double
Gaussian state (\ref{eq:state2G}), although calculations are more
involved. A simple application of (\ref{eq:negativity}) to Eq. (\ref{eq:WFiterated})
leads to the result 
\begin{equation}
\eta(t)=(1-p)^{t}
\end{equation}
for the negativity as a function of time. This simple result can be
interpreted as the damping of the out-of-diagonal terms in (\ref{eq:WFiterated}).
As time goes on, these interference terms tend to fade away, and one
is left with an incoherent state with a positive Wigner function.
This transition from a coherent superposition to an incoherent one
is, of course, a well known phenomenon in the theory of open quantum
systems which shows a change in the nature of the Wigner function
that is monitored by our definition of the corresponding negativity.

Although it is obvious from this discussion that in the presence of
spin the negativity does not have the clear unique physical meaning
it had in the purely spatial case (either continuous or discrete),
the quantity $\eta$ introduced here may be useful to characterize
some features of the quantum state or the dynamics when the study
is restricted to particular families of states. The topic is nevertheless
far from being closed, and could be the subject of further debate.

\section{Conclusions}

In this paper, we have elaborated the previously introduced Wigner
formalism for a particle in an infinite 1D lattice, in order to account
for dynamics and for the presence of an additional, finite-dimensional,
degree of freedom. Our goal was to describe the dynamics on the phase
space associated to this problem. Although we have concentrated, for
simplicity, on the case where such additional degree of freedom corresponds
to a spin 1/2, one can envisage more general situations where higher
spins, or different properties, such as the polarization of a photon,
are considered. As we have showed, the matrix formalism is specially
well suited to describe the interaction of the particle with a spin-dependent
Hamiltonian on a fixed basis, and keeps a close resemblance to the
relativistic WF formalism \cite{PhysRevD.13.950,Hakim:1379544}, a
fact that might be useful in the investigation of the non relativistic
limit of a given problem. We have illustrated the construction of
the WF by analyzing first some simple static examples, like the ``Schrödinger
cat'' double delta or two-Gaussian states. For these states, the
position and spin variables are entangled, and this entanglement manifests
in a particular structure of the WM.

We have studied the time evolution of the WM for some simple cases.
We have explicitly shown the equation governing the evolution of the
WF for a general space-dependent potential. This equation, however,
can only be exactly solved for some special cases, as we have done
for the case of a linear potential, where one recovers the well known
phenomenon of Bloch oscillations. A similar statement is valid for
a Hamiltonian that can be factored as a scalar part and a spin operator.
We have obtained the equation of motion for a general scalar term,
and solved it in the linear case, what allows us to compare with the
dynamics in the spinless case. The presence of a ``spin dependent
force'' introduces new features on the dynamics that manifest in
phase space. To complete the above description, we have incorporated
the role of decoherence which, for some simple examples, can be implemented
for the WM in a closed form.

In some physical situations, the interaction appears as short pulses
acting on the particle, a paradigmatic example being the Quantum Walk.
It is possible to analyze the role of decoherence also in this case,
and we have analyzed a simple example for the double delta state,
when decoherence appears as projections either on the spatial or in
the original spin basis. We have showed that both kind of mechanisms
produce the same effect, which translates into a damping of the off-diagonal
matrix components.

Finally, we have explored a possible extension of the concept of negativity,
as defined for the scalar WF, to the spin 1/2 case. While it is not
evident what the physical meaning of such negativity might have once
the spin is incorporated to the particle, we have proposed the minimum
requirements that, in our opinion, this magnitude should obey, and
we have suggested a definition of negativity that fulfills these requirements.
Following this proposal, we analyzed how decoherence translates into
a decreasing of negativity in the above decoherence model. We also
showed that our definition of negativity has not trivial correspondence
with entanglement, as clearly indicated by an analysis of the Werner
state. We think, however, that it is worth studying further the relationship
of the Wigner description to the quantum properties of general states
in a lattice.
\begin{acknowledgments}
This work has been supported by the Spanish Ministerio de Educación
e Innovación, MICIN-FEDER project FPA2011-23897 and ``Generalitat
Valenciana'' grant GVPROMETEOII2014-087. M.H. and A.P. would like
to express their gratitude for the hospitality of J. I. Cirac and
MPQ at Garching, where part of this research was developed, and to
\textit{Centro de Ciencias Pedro Pascual} in Benasque, Spain.
\end{acknowledgments}

\section{Appendix A. Dynamics of the Wigner function on a lattice for a particle
subject to a potential}

We will derive the differential equation that is obeyed by the WM
in two cases: I) A particle interacting with a position-dependent
potential $V(x)$ and II) A spin 1/2 particle under the effect of
a spin-position Hamiltonian of the form (\ref{eq:Hlatticespin}).

I) We start with the Hamiltonian defined in (\ref{eq:Hlattice}).
The interaction in this case only affects the phase space variables
$(m,k)$, therefore spin indices can be omitted for the moment, but
can be recovered in the final expression by replacing $W(m,k,t)\longrightarrow W_{\alpha\beta}(m,k,t)$.
Of course, for a spinless particle no replacement is necessary. 

The evolution equation is obtained from the von Neumann equation for
the density operator (\ref{eq:von Neumann}). Using the properties
of the $T_{\pm}$ operators one obtains 
\begin{equation}
\frac{\partial}{\partial t}W(m,k,t)=2J\sin k\left[W(m+1,k,t)-W(m-1,k,t)\right]+D,\label{eq:WnospintrhoD}
\end{equation}

where 
\begin{equation}
D\equiv-\frac{i}{2\pi}\sum_{l}e^{-i(2l-m)k}(V_{l}-V_{m-l})\braket{l|\rho(t)|m-l}.
\end{equation}
We assume that $V(x)$ is continuous and infinitely derivable at any
point. Remembering that $V_{l}=V(la)$, we Taylor expand both $V_{l}$
and $V_{m-l}$ around the point $\frac{m}{2}a$, so that
\begin{equation}
D=-\frac{i}{2\pi}\sum_{l}e^{-i(2l-m)k}\sum_{p=0}^{\infty}\frac{a^{p}}{p!}\left.\frac{d^{p}V(x)}{dx^{p}}\right|{}_{x=ma/2}\frac{(2l-m)^{p}}{2^{p}}[1-(-1)^{p}]\braket{l|\rho(t)|m-l}.
\end{equation}
With the help of the WF definition, Eq. (\ref{eq:WFspin}), one arrives
to 
\begin{equation}
D=-i\sum_{p=0}^{\infty}\frac{a^{p}}{p!}\frac{d^{p}V(x)}{dx^{p}}\Bigg|{}_{x=ma/2}\frac{1}{(-2i)^{p}}[1-(-1)^{p}]\frac{\partial^{p}W(m,k,t)}{\partial k^{p}}.
\end{equation}
Notice that even values of $p$ do not contribute in the above sum,
so we restrict ourselves to odd values $p=2s+1$ with $s\in\mathbb{N}$.
After simplifying, we finally obtain 
\begin{equation}
\frac{\partial}{\partial t}W(m,k,t)=2J\sin k\left[W(m+1,k,t)-W(m-1,k,t)\right]+\sum_{s=0}^{\infty}\frac{(-1)^{s}a^{2s+1}}{2^{2s}(2s+1)!}\left.\frac{d^{2s+1}V(x)}{dx^{2s+1}}\right|{}_{x=ma/2}\frac{\partial^{2s+1}W(m,k,t)}{\partial k^{2s+1}}.\label{eq:Wnospintser-1}
\end{equation}

II) We now develop an equation of motion for a spin 1/2 particle which
is subject to a spin position-dependent interaction given by Eq. (\ref{eq:Hlatticespin}).
Following similar steps to the previous case, and making use of $\sigma_{z}\ket{\alpha}=(-1)^{\alpha}\ket{\alpha}$,
$\alpha=0,1$, one gets

\begin{equation}
\frac{\partial}{\partial t}W_{\alpha\beta}(m,k,t)=2J\sin k\left[W_{\alpha\beta}(m+1,k,t)-W_{\alpha\beta}(m-1,k,t)\right]+D_{\alpha\beta},\label{eq:WnospintrhoDspin}
\end{equation}
with 
\begin{equation}
D_{\alpha\beta}\equiv-\frac{i}{2\pi}\sum_{l}e^{-i(2l-m)k}\left[(-1)^{\alpha}V_{l}-(-1)^{\beta}V_{m-l}\right]\langle n,\alpha|\rho|m-n,\beta\rangle.
\end{equation}
After expanding $V_{l}$ and $V_{m-l}$ around the point $\frac{m}{2}a$
as before, we arrive to
\begin{equation}
D_{\alpha\beta}=-\frac{i}{2\pi}\sum_{l}e^{-i(2l-m)k}\sum_{p=0}^{\infty}\frac{a^{p}}{p!}\left.\frac{d^{p}V(x)}{dx^{p}}\right|{}_{x=ma/2}\frac{(2l-m)^{p}}{2^{p}}(-1)^{\alpha}[1-(-1)^{p}(-1)^{\alpha+\beta}]\braket{l|\rho(t)|m-l}.
\end{equation}
In terms of the WM, 
\begin{equation}
D_{\alpha\beta}=-i\sum_{p=0}^{\infty}\frac{a^{p}}{p!}\left.\frac{d^{p}V(x)}{dx^{p}}\right|{}_{x=ma/2}\frac{1}{(-2i)^{p}}(-1)^{\alpha}[1-(-1)^{p}(-1)^{\alpha+\beta}]\frac{\partial^{p}W_{\alpha\beta}(m,k,t)}{\partial k^{p}}.
\end{equation}
In order to determine the values of $p$ that contribute to the above
sum, one has to consider two different cases. 

If $\alpha=\beta$, only odd values $p=2s+1$ with $s\in\mathbb{N}$
have to be considered, and one is lead to
\begin{equation}
D_{\alpha\alpha}=(-1)^{\alpha}\sum_{s=0}^{\infty}\frac{(-1)^{s}a^{2s+1}}{(2s+1)!}\frac{1}{2^{2s}}\left.\frac{d^{2s+1}V(x)}{dx^{2s+1}}\right|{}_{x=ma/2}\frac{\partial^{2s+1}W_{\alpha\alpha}(m,k,t)}{\partial k^{2s+1}},
\end{equation}
whereas for the off-diagonal elements $\alpha\neq\beta$ we have now
only the contribution from even values of $p=2s$, and we can easily
obtain 
\begin{equation}
D_{\alpha\beta}=-2i(-1)^{\alpha}\sum_{s=0}^{\infty}\frac{(-1)^{s}a^{2s}}{(2s)!}\frac{1}{2^{2s}}\left.\frac{d^{2s}V(x)}{dx^{2s}}\right|{}_{x=ma/2}\frac{\partial^{2s}W_{\alpha\beta}(m,k,t)}{\partial k^{2s}}.
\end{equation}
\bibliographystyle{iop}
\bibliography{books,/home/perez/Dropbox/biblio/qwalks,/home/perez/Dropbox/biblio/wigner}

\end{document}